# Controlling Electromagnetic Fields at Boundaries of Arbitrary Geometries

Jonathon Yi Han Teo[1], Liang Jie Wong[1], Carlo Molardi[1], and Patrice Genevet[1,2*]

[1] Singapore Institute of Manufacturing Technology, Agency for Science, Technology and Research (A*STAR), 71 Nanyang Drive, Singapore, 638075, Singapore

[2] Centre de Recherche sur l'Hétéro-Epitaxie et ses Application (CRHEA, UPR 10), CNRS, Rue Bernard Gregory, Sophia-Antipolis, 06560 Valbonne, France.

Corresponding author: genevetp@simtech.a-star.edu.sg

**Rapid developments in the emerging field of stretchable and conformable photonics necessitate analytical expressions for boundary conditions at metasurfaces of arbitrary geometries. Here, we introduce the concept of *conformal boundary optics*: a design theory that determines the optical response for designer input and output fields at such interfaces. Given any object, we can realise coatings to achieve exotic effects like optical illusions and anomalous diffraction behaviour. This approach is relevant to a broad range of applications from conventional refractive optics to the design of the next-generation of wearable optical components. This concept can be generalized to other fields of research where designer interfaces with nontrivial geometries are encountered.**

In a bulk medium, a wave (e.g., optical, sound, seismic) accumulates phase gradually and propagates without experiencing abrupt variations. At the boundary with another material, however, the wave can experience large -although physically admissible- discontinuities in its reflected and transmitted fields [1,2], as dictated by the boundary conditions of the system. The behavior of optical waves at interfaces plays a leading role in many industries: for example, in the hydrocarbon industry, where reverse seismic refraction is used to map petroleum reserves deep inside the soil of the planet [3]. In optics, reflection and refraction at interfaces are central to the design of optical components (e.g., mirrors, windows, waveplates, and lenses) [4]. Recently, several unexpected interfacial optical effects of practical interests have been demonstrated. These include anomalous reflection and refraction [5-10], giant spin-Hall effect [11], flat lensing [12], controlled Cherenkov surface plasmon emission [13], holography [14-16] and surface cloaking [17].

Here, we introduce the concept of conformal boundary optics, an analytical method – based on novel, first-principle derivations – that allows us to engineer transmission ($E^t$) and reflection ($E^{ref}$) at will for any interface geometry and any given incident wave ($E^{inc}$). By resolving the boundary conditions between two materials at an interface of arbitrary geometry, this method addresses recent developments in nanophotonics with the general technique of differential geometry and coordinates transformation. Unlike transformation

(1)

optics, our approach deals directly with abrupt changes in the fields, and therefore acts at the level of the boundary conditions of the electromagnetic fields. Whereas transformation optics determines bulk optical properties by exploiting the relationship between a given coordinate system and the coordinate system that conforms to the travel of light [18-28], the proposed concept determines the optical properties of a *metasurface of arbitrary geometry* by exploiting the relationship between a given ambient coordinate system and the coordinate system that conforms *to the geometry of the boundary*. While a powerful concept in itself, the mathematical derivation associated with its analytical formulation – which we present in detail in the supplementary information – is highly non-trivial since it cannot be generalized from existing boundary conditions for generic surface geometries. This concept provides a wide range of new design opportunities, for example, to hide objects behind an "optical curtain", to create optical illusions by reflecting virtual images, or to suppress the diffraction generally occurring during light scattering at corrugated interfaces (Fig. 1**a**).

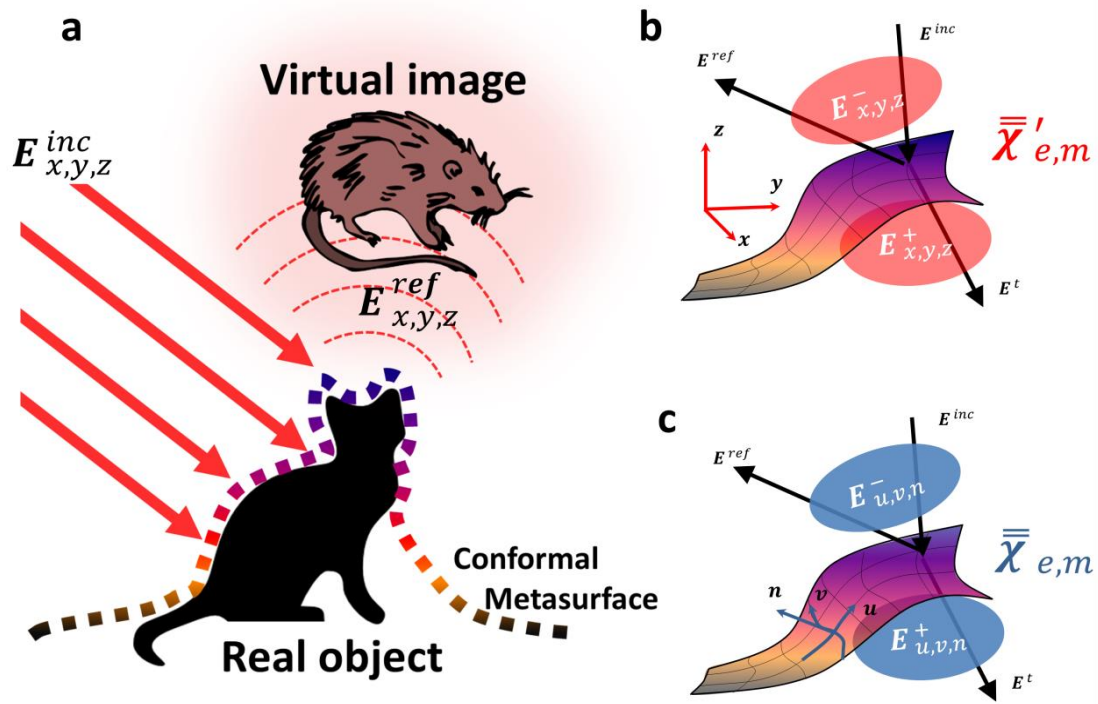

**Figure 1| The concept of conformal boundary optics.** **The surface susceptibility of an interface that conforms to an object can be designed to produce an arbitrary, user-specified response. In a, a cat is visually transmogrified into a rat as an incident light $E^{inc}_{x,y,z}$ is converted at the object interface to produce a user-specified virtual holographic image $E^{ref}_{x,y,z}$ in the far-field. This way, the actual geometry of the object no longer restricts the properties of the reflected light. Our proposed concept exploits the relationship between a given ambient coordinate system (b) and the coordinate system that conforms to the geometry of the boundary (c). $\bar{\bar{\chi}}_{e,m}$ and $\bar{\bar{\chi}}'_{e,m}$ denote the optical response of the interface – quantities called surface susceptibilities tensors – in these coordinate systems respectively.**

(2)

The transmission and reflection of waves with ultrathin interfaces ($\delta \ll \lambda$) has been recently demonstrated in various experiments reviewed by refs. [6,9,10]. Abrupt modifications of the fields across an interface are engineered by depositing an array of sub-wavelength resonators specifically tailored to address local amplitude, phase and polarization changes in the light traversing the interface. Several examples have been reported on the control of the radiation patterns of thermal, acoustic, seismic and electromagnetic waves, indicating that these are probably the most suitable tools for engineering the discontinuities at will. This technique – which has been dubbed "metasurface physics" and has given rise to a broad range of applications across physics, chemistry, biology and materials science – is considered to be one of the most promising and disruptive emerging technologies of recent times. The concept of conformal boundary optics pushes the limits of metasurface physics beyond the design of simple planar interfaces. This has fascinating applications, accounting for phenomena as diverse as the engineering of modal field distributions of optical resonators with various shapes, and the creation of "optical illusions", which are required in cloaking, virtual imaging, and kinoform holography .

It is well-understood that the far-field image of an object is the result of light reflected or transmitted from its surface. By choosing the appropriate boundary conditions, therefore, one can modify the reflection and transmission of light to produce various kinds of unexpected optical effects. For user-specified incident, transmitted and reflected fields, and a given surface geometry represented by coordinate system $(u, v, n)$, one can determine the required surface susceptibility tensors $\bar{\bar{\chi}}_{e,m}$ using our concept. This technique accommodates any choice of ambient coordinate system, which happens to be Cartesian in Fig. 1**b** and 1**c** where the electromagnetic fields before and after the interface are denoted $E_{x,y,z}^{\mp}$. Notably, the technique yields surface susceptibilities in the coordinate system of the interface, which is necessary when the optical components are manufactured.

**Generalized sheet transition conditions in Cartesian coordinates**

To date, a rigorous expression of the electromagnetic boundary conditions at designer interfaces has been proposed only for planar interfaces where the electromagnetic fields are defined using a Cartesian coordinate system [31-41]. These equations are known as the generalized sheet transition conditions (GSTCs). From a physical point of view, discontinuities in electromagnetic fields across any regular surface depend upon the constitutive parameters of the interface: namely, surface charge density $\rho$ , the current density $j$, the induced dipole moments at the interface and the optical response of the surrounding media. This requires ($\rho$, $j$) and the fields $E, H, p$ and $m$ in the Maxwell's equations to be expressed in the sense of distributions, where $E, H$ are respectively the electric and magnetic fields, and $p$ and $m$ respectively represent the surface electric and magnetic induced currents derived by averaging the local fields of the electric and magnetic induced dipole moments in the plane z=0. In a planar configuration, writing each variable as $\zeta(z) = \{\zeta(z)\} + \sum_{k=0}^{N} \zeta_k \delta^{(k)}(z)$ , with the function $\zeta(z)$ discontinuous at $z = 0$, it is possible to derive a set of generalized sheet transition conditions (GSTCs) for the electromagnetic fields [31-33].

**Expression of the generalized sheet boundary conditions in local coordinate**

Obtaining the electromagnetic boundary conditions at a nonplanar interface (Fig. 2) is highly non-trivial because the GSTCs apply only to interfaces whose local coordinate systems are Cartesian.

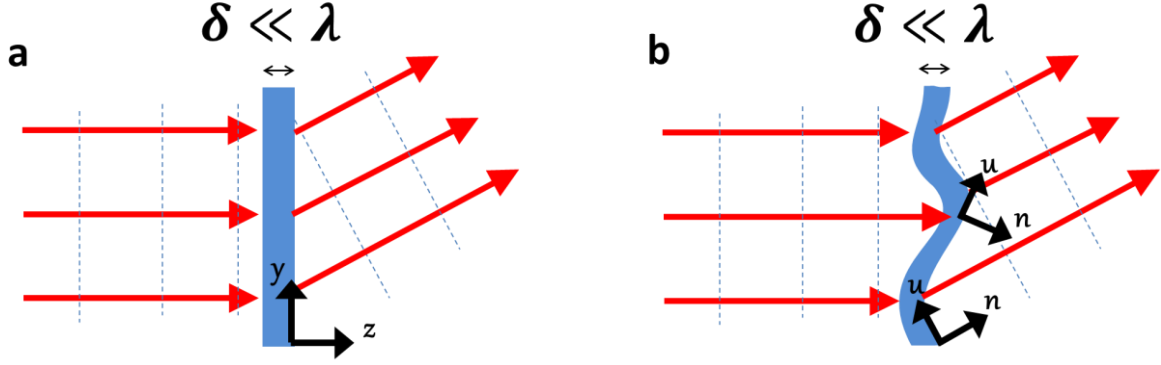

**Figure 2| The challenge of designing non-planar metasurfaces. a),** a 2D planar metasurface of sub-wavelength thickness $\delta \ll \lambda$, can transmit any incident optical field at a specific angle by imposing a gradient of phase discontinuity. For planar interfaces, GSTC boundary conditions readily apply and the surface susceptibility tensors can be calculated. **b,** The local coordinate system of the surface follows its local curvature, and therefore it changes with the position along the interface. Boundary conditions of the fields are obtained in the coordinate system of the interface, and are therefore position dependent. To produce an effect equivalent to that in b, the surface susceptibilities of the optical interface have to be engineered to account for the effect of the physical distortion. The dashed blue lines denote the equiphase fronts of the electromagnetic fields.

One cannot generalize the existing GSTC boundary expressions to a non-Cartesian coordinate system because it is unclear if the components are covariant or contravariant. In the Cartesian coordinate system of GSTCs, the tangent-cotangent isomorphism is the identity, meaning that there is no difference between the covariant and contravariant components. This is not true for coordinate systems in general. Considering a volumetric boundary of sub-wavelength thickness – as elucidated in the supplementary materials -- we treat $S$ as an interface and define $(u, v)$ to be the coordinate system that conforms to it. This approximation is valid for optical metasurfaces of subwavelength thicknesses. As such, the treatment of general, non-planar surfaces necessitates novel first-principle derivations starting from the integral forms of Maxwell's equations. The required surface susceptibility tensors for a metasurface of arbitrary geometry and user-specified functionality are

$$\chi_{e,m}^{ik'} = \begin{cases} \sqrt{g^S}\chi_{e,m}^{ik}\Lambda_k^{k'} & for\ i = u, v\ and\ k' = x, y, z, \\ \chi_{e,m}^{ik}\Lambda_k^{k'} & for\ i = n\ and\ k' = x, y, z. \end{cases} \qquad (1)$$

where $\bar{\bar{\chi}}_{e,m}$ are the desired surface susceptibility tensors expressed in the local coordinate system. $\bar{\bar{\chi}}'_{e,m}$ is the expression of $\bar{\bar{\chi}}_{e,m}$ in the ambient coordinates (typically Cartesian) of the system. The $\Lambda_i^{i'}$ is the transformation matrix used to change the components of the covector field, given by $V_i = \Lambda_i^{i'} V_{i'}$, $g^S$ is a normalization factor related to the surface geometry, and $\bar{\bar{\chi}}_{e,m}$ are related to the user-specified input fields ($E_{x,y,z}^{inc}$ and $H_{x,y,z}^{inc}$), output fields ($E_{x,y,z}^{ref,t}$ and $H_{x,y,z}^{ref,t}$) and the geometry of the surface by the electromagnetic boundary conditions:

$$[ij]\, \Lambda_j^{k'} E_{k'}\big|_{-}^{+} e_i = [ij]\partial_j\big(\chi_e^{nk'} E_{k'}^{\overline{av}}\big) e_i + \partial_t\big(\chi_m^{ik'} B_{k'}^{\overline{av}}\big) e_i \qquad (2.a)$$

$$[ij]\, \Lambda_j^{k'} H_{k'}\big|_{-}^{+} e_i = [ij]\partial_j\big(\chi_m^{nk'} H_{k'}^{\overline{av}}\big) e_i - \partial_t\big(\chi_e^{ik'} D_{k'}^{\overline{av}}\big) e_i \qquad (2.b)$$

$$\Lambda_n^{k'} D_k\big|_{-}^{+} + \sqrt{g^S}\partial_i\big(\chi_e^{ik'} D_{k'}^{\overline{av}}\big) = 0 \qquad (2.c)$$

$$\Lambda_n^{k'} B_{k'}\big|_{-}^{+} + \sqrt{g^S}\partial_i\big(\chi_m^{ik'} B_{k'}^{\overline{av}}\big) = 0 \qquad (2.d)$$

where $i, j = u, v$ and $k = u, v, n$ and $[ij] = \begin{cases} 1 & \text{if } i = u \text{ and } j = v, \\ -1 & \text{if } i = v \text{ and } j = u, \\ 0 & \text{otherwise.} \end{cases}$ \qquad (2.e)

$\partial_j$ and $\partial_t$ denote differentiation with respect to the coordinate $j$, and differentiation with respect to time respectively. In the expressions above, for any field $\mathbf{V}$, $V_j|_{-}^{+}$ and $V_j^{\overline{av}}$ respectively represent the difference in the $j$-th component of the fields on either side of the metasurface ($V_j^- = V_j^{inc} + V_j^{ref}$ and $V_j^+ = V_j^t$), and the $j$-th component of the average of the fields. For a flat surface (x=u, y=v z=0, $\Lambda_j^{i'} = \delta_j^{j'}$, and $\sqrt{g^S} = 1$), reduces to the GSTCs derived in [36] (see additional material for details of the full derivation).

The set of equations (1 and 2) is thus a powerful means of obtaining the analytical susceptibilities of metasurfaces that are non-planar or are required to be wearable, conformable and/or stretchable. To the best of our knowledge, the findings summarized in (2a-d) represent the first time that electromagnetic boundary conditions have been obtained for interfaces of arbitrary geometries. As we shall demonstrate, our findings also have important implications for understanding the behaviour of light at regular interfaces between two dielectrics. In the following section, we illustrate the versatility of this technique with two examples. In the first example, we analytically obtain the surface susceptibility of a mantle cloak. In the second example, we analytically obtain surface susceptibilities that suppress the effect of diffraction at corrugated interfaces, thereby redirecting light along unconventional directions.

**Optical illusions and anomalous diffraction at conformable metasurfaces**

The prospect of cloaking has intrigued the scientific community and drawn considerable attention over the last 10 years. Here, we show that our technique allows one to analytically obtain the surface properties of a mantle cloak in the reflection mode [17]. Suppose that the object surface function can be written as $S = n^{-1}(0)$, where $n(x,y,z) = z - \cos(x) - \cos(y)$ and S is the graph of $f(u,v) = \cos u + \cos v$ (Fig. 3**a** and supplementary material), we calculate the surface susceptibility that matches an incident plane wave

(5)

propagating at angle $\theta$ to a reflected plane wave propagating at the same angle $\theta$ for all orientations of the surface. This interface produces an equivalent effect of a conventional planar mirror. Setting the transmission to zero, we can convert an incident TE field, $E_y^{inc} = A_{TE}\xi^{inc}$ to $E_y^{ref} = A_{TE}\xi^{ref}$ where $\xi^{inc} = e^{i(\omega t - k_0(x\sin\theta + z\cos\theta))}$ and $\xi^{ref} = e^{i(\omega t - k_0(x\sin\theta - z\cos\theta))}$ with the following susceptibilities (which apply for an incident TM mode as well):

$$\chi_{e,m}^{uu} = \frac{2ic}{\omega\gamma} \left( \frac{1+\sin^2 v}{\cos\theta} \frac{\xi^{inc}+\xi^{ref}}{\xi^{ref}-\xi^{inc}} - \sin\theta \sin u \sin^2 v \right) \tag{3a}$$

$$\chi_{e,m}^{uv} = \frac{2ic}{\omega\gamma} \sin v \left( -\frac{\sin u}{\cos\theta} \frac{\xi^{inc}+\xi^{ref}}{\xi^{ref}-\xi^{inc}} + \sin\theta (1+\sin^2 u) \right) \tag{3b}$$

$$\chi_{e,m}^{un} = \frac{2ic}{|\nabla n|^2 \omega} \left( \frac{\sin v}{\cos\theta} \frac{\xi^{inc}+\xi^{ref}}{\xi^{ref}-\xi^{inc}} + \sin\theta \sin^2 v \right) \tag{3c}$$

$$\chi_{e,m}^{vu} = \frac{2ic}{\omega\gamma} \sin v \sin u \left( -\cos\theta \frac{\xi^{ref}-\xi^{inc}}{\xi^{ref}+\xi^{inc}} + \sin u \sin\theta \right) \tag{3d}$$

$$\chi_{e,m}^{vv} = \frac{2ic}{\omega\gamma} (1+\sin^2 u) \left( \cos\theta \frac{\xi^{ref}-\xi^{inc}}{\xi^{ref}+\xi^{inc}} - \sin u \sin\theta \right) \tag{3e}$$

$$\chi_{e,m}^{vn} = \frac{2ic}{|\nabla n|^2 \omega} \left( \cos\theta \frac{\xi^{ref}-\xi^{inc}}{\xi^{ref}+\xi^{inc}} - \sin u \sin\theta \right) \tag{3f}$$

Where $|\nabla n|^2 = 1 + \sin^2 u + \sin^2 v$ and $\gamma = [1 + \sin^2 u + \sin^2 v]^{3/2}$. Note that the susceptibility in (3) is a function of incidence angle, implying that an omni-directional carpet cloak cannot be designed with only linear susceptibilities. The general expression for $\bar{\bar{\chi}}_{e,m}$ can be used, for example, to create a virtual image at the detector plane. Details of this methodology are provided in section 3.1 of the supplementary material.

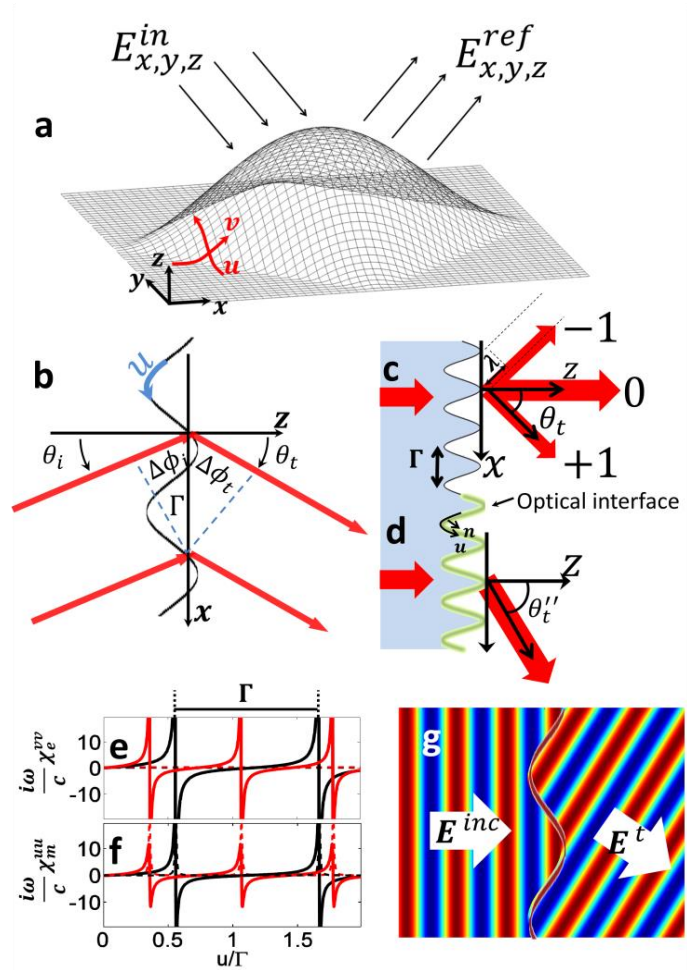

**Figure 3. Designing cloaks and anomalous diffraction with conformal boundary optics.** Mirror surface cloak described by $f(u,v) = \cos u + \cos v$ reflects an incident plane wave at an angle of $\theta$ with respect to the z-axis to form a plane wave propagating at the same angle $\theta$ for all orientation of the surface. b, the relation between the angle of the incoming and outgoing waves can be imposed at the interface by considering the condition that the difference between the propagation phase shift of two light rays of the incident wavefront ($\Delta\phi_i$), impinging on the surface at points separated by a distance ($\Delta u, \Delta v$), and the propagation phase shift after the interface ($\Delta\phi_t$) is exactly compensated by the phase shift introduced at the interface at those points. However, local phase retardations are not the only physical quantities to take into consideration. Because the orientation of the surface– specifically, the orientation of the unit vector field $N$ normal to $S$ – changes as one moves across it, the induced dipole moments at the surface are excited and radiate differently depending on the location. To account for the aforementioned changes in phase, amplitude and polarization, susceptibility tensors are calculated from Eq. (1) and (2). c and d respectively show a conventional and an anomalous grating. Light incident on a metasurface-grating (d) can blaze the diffraction towards a single order (black curves in e and f) but it can also refract light at any other user-specified angle (red curves in e and f). Solid (dashed) curves represent the imaginary (real) part of the susceptibility tensors. g shows the calculation of the real part of the electric fields that satisfy the conditions

(7)

in Eq.5 for free-standing metasurfaces in air, that is, when the refractive indices on either side of the boundary are set to be equal.

In our second example, we design a reflection-less optical interface [42] defined by a periodic structure of period $\Gamma$ greater than the wavelength of the light, $\Gamma > \lambda_0$ with a surface function given by $f(u,v) = \sin\frac{2\pi}{\Gamma}u$. Considering light incident at an angle $\theta_i$ from a lossless dielectric of refractive index $n_i$, as in Fig. 3**b**, conventional diffraction in Fig 3.**c** creates plane waves at angles $\theta_t$ given by the grating formula:

$$m\lambda_0 = \Gamma(n_i \sin\theta_i - \sin\theta_t) \tag{4}$$

By depositing a metasurface conformable to the periodically undulating interface, one can tailor the diffracted light. As shown in Fig. 3**b**, the interface has to delay the incoming fields such that the transmitted light from each point along the interface constructively interferes along the user preferred direction, creating plane waves that travel along the angle $\theta_t$. Suppose a TE electromagnetic field normally incident on the metasurface $E_y^{inc} = A_{TE}\xi^{inc}$ where $\xi^{inc} = e^{i(\omega t - k_0 z)}$. Our objective is to calculate the interface properties to produce a transmitted field $A_{TE}\xi^t$ where $\xi^t = e^{i(\omega t - k_0(x\sin\theta_t + z\cos\theta_t))}$ (where $\theta_t$ is no longer restricted by the grating law in Eq.4). After solving (2), we obtain the susceptibility tensors as

$$\chi_m^{uu} = \frac{2c}{i\omega\left(1 + \frac{4\pi^2}{\Gamma^2}\cos^2\frac{2\pi u}{\Gamma}\right)^{3/2}} \left(\frac{\xi^{inc} - \xi^t}{\xi^{inc} + \xi^t\cos\theta_t}\right) \tag{5a}$$

$$\chi_m^{un} = \frac{2c\,\cos u}{i\omega(1 + \frac{4\pi^2}{\Gamma^2}\cos^2\frac{2\pi u}{\Gamma})} \left(\frac{\xi^{inc} - \xi^t}{\xi^{inc} + \xi^t\cos\theta_t}\right) \tag{5b}$$

$$\chi_m^{vn} = \frac{1}{1 + \frac{4\pi^2}{\Gamma^2}\cos^2\frac{2\pi u}{\Gamma}} \left(\frac{\xi^+ \sin\theta_t}{\sqrt{1 + \frac{4\pi^2}{\Gamma^2}\cos^2\frac{2\pi u}{\Gamma}}\,(\xi^{inc} + \xi^t\cos\theta_t)} - 1\right) \tag{5c}$$

$$\chi_e^{vv} = \frac{2c}{i\omega\sqrt{1 + \frac{4\pi^2}{\Gamma^2}\cos^2\frac{2\pi u}{\Gamma}}} \left(\frac{\xi^{inc} - \xi^t\left(\frac{2\pi}{\Gamma}\cos\frac{2\pi u}{\Gamma}\sin\theta_t - \cos\theta_t\right)}{\xi^{inc} + \xi^t}\right) \tag{5d}$$

All of the other components are equal to zero. In the supplementary material, we perform further analytical calculations to verify that the transmission refracts at arbitrary angle $\theta_t$.

Our use of coordinate transformations to control abrupt changes of light at interfaces is complementary to – and distinct from -- the technique of transformation optics, which uses coordinate transformations to control the propagation of light in bulk media. Whereas transformation optics uses a coordinate system that conforms to the direction of light propagation, the concept of conformal boundary optics uses a coordinate system that conforms to the geometry of the interface. With the advent of transformation optics and metamaterials, many creative ways to manipulate light have been proposed that involve negative refraction, hyperlensing, cloaking and backward Cherenkov emission [43-47]. Transformation optics has also been used to redirect the propagation of surface electromagnetic waves [25-28], and to suppress the surface wave scattering losses [48].

New and interesting optical effects are expected with the added control of fields at interfaces of metamaterials with arbitrary geometries (Fig. 4).

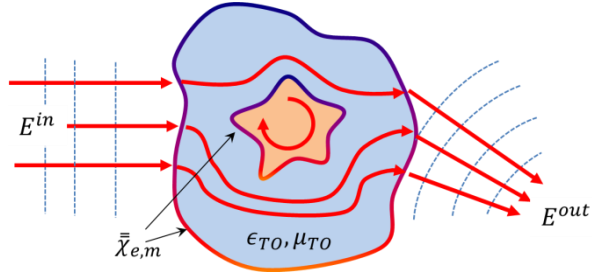

**Figure 4. <u>Comprehensive control of electromagnetic fields with metasurfaces and metamaterials.</u> We envision combining the use of conformal boundary optics and transformation optics to achieve complete, geometry-independent control over the behavior of light at interfaces and in bulk media.**

The concept of conformal boundary optics has important implications even in the realm of conventional optics, where the surface susceptibility tensors are zero. (2) may then be simplified to

$$[ij]\,\Lambda_j^{k'} E_{k'}|_-^+ \, \boldsymbol{e}_i = 0 \tag{6a}$$

$$[ij]\,\Lambda_j^{k'} H_{k'}|_-^+ \, \boldsymbol{e}_i = J_S^i \, \boldsymbol{e}_i \tag{6b}$$

$$\Lambda_n^{k'} D_{k'}\Big|_-^+ = \rho_s \tag{6c}$$

$$\Lambda_n^{k'} B_{k'}\Big|_-^+ = 0 \tag{6d}$$

In many optics textbooks, the boundary conditions are derived by considering an Amperian loop and a Gaussian pillbox at an interface, often graphically represented -- for aesthetical reasons -- as a nonplanar surface. This, however, can be misleading because these conditions are originally derived in the coordinate system that conforms to the interface whereas the electromagnetic fields are expressed in the ambient -- typically Cartesian, spherical or cylindrical -- coordinate system. Because the surface normal of the nonplanar interface is changing, the boundary conditions vary depending on the position along the interface. This observation shows that our approach is already significant for conventional interfaces that involve only regular dielectric materials. Prior to this work, boundary conditions existed only for cases where the interface -- which the coordinate system conforms to – has a basic geometry e.g., Cartesian, cylindrical, spherical. We believe that the approach discussed in this paper constitutes a first step towards the development of a general theory for waves in systems containing interfaces with complex geometries.

**Conclusion**

We have introduced the concept of conformal boundary optics, a versatile design theory that enables metasurfaces of arbitrary geometries to find even greater applications in the emerging fields of stretchable, camouflage photonics and augmented reality. One can envision, for instance, changing the appearance of buildings or any kind of objects. Controlling light at curved geometries will find widespread industrial applications, for example, in the design of the next generation of contact lenses and flexible, stretchable optical surfaces. One can also realise unusually-shaped optical cavities that support designer resonant modes. By allowing us to transcend the limitations imposed by physical geometries, this concept has fundamental implications for the way in which we understand and structure light in systems of arbitrary geometries.

**References and Notes**

1. M. Born and E. Wolf (1999), *Principles of optics*, 7$^{th}$ Edition
2. Griffiths, David J. (2007) *Introduction to Electrodynamics*, 3rd Edition
3. Dobrin, Milton B. and Savit, Carl H. (1988) *Introduction to Geophysical Prospecting*, 4th Edition
4. Kingslake R. and R.B. Johnson, *Lens Design Fundamentals*, 2$^{nd}$ ed (Academic,2009)
5. Yu, N., Genevet, P., Kats, M. A., Aieta, F., Tetienne, J. P., Capasso, F., & Gaburro, Z. (2011). *Light propagation with phase discontinuities: generalized laws of reflection and refraction. Science*, 334(6054), 333-337.
6. AV Kildishev, A Boltasseva, VM Shalaev, *Planar Photonics with Metasurfaces*, Science 339, 6125 (2013).
7. Yong Li, Bin Liang, Zhong-ming Gu, Xin-ye Zou and Jian-chun Cheng, *Reflected wavefront manipulation based on ultrathin planar acoustic metasurfaces*, Scientific Reports 3, 2546 (2013)
8. S. Brûlé, E. H. Javelaud, S. Enoch, and S. Guenneau, *Experiments on Seismic Metamaterials: Molding Surface Waves*, Physical Review Letters 112, 133901 (2014)
9. Genevet, P. and Capasso, F*., Flat Optics: Wavefronts Control With Huygens' Interfaces*, IEEE Photonics 6, 0700404 (2014).
10. Yu, N., & Capasso, F. (2014). *Flat optics with designer metasurfaces*. Nature materials, 13(2), 139-150.
11. Xiaobo Yin, Ziliang Ye, Junsuk Rho, Yuan Wang, Xiang Zhang, *Photonic Spin Hall Effect at Metasurfaces,* Science 339, 1405 (2013)
12. Aieta, F., Genevet, P., Kats, M. A., Yu, N., Blanchard, R., Gaburro, Z., & Capasso, F. (2012). *Aberration-free ultrathin at lenses and axicons at telecom wavelengths based on plasmonic metasurfaces*. Nano letters, 12(9), 4932-4936.
13. Genevet, P., Wintz, D., Ambrosio, A., She, A., Blanchard, R., Capasso, F. , *Controlled steering of Cherenkov surface plasmon wakes with a one-dimensional metamaterial*, Nature Nano., doi:10.1038/nnano.2015.137 (2015)
14. Larouche et al, Infrared *Metamaterial phase hologram,* Nature Mat. 11, 450 (2012)
15. Zheng, G. et al., *Metasurface holograms reaching 80% efficiency*, Nature Nano., 10, 308 (2015).
16. Genevet P. and F. Capasso, *Holographic optical metasurfaces: a review of current progress*, Rep. Prog. Phys. 78 (2015) 024401.
17. Mohammadi Estakhri, N., & Alu, A. *Ultrathin Unidirectional Carpet Cloak and Wave-front Reconstruction with Graded Metasurfaces*. Antennas and Wireless Propagation Letters, IEEE Vol.13, Page 1775-1778 (2014).
18. Pendry, J. B., Schurig, D., & Smith, D. R. (2006). *Controlling electromagnetic fields*. science, 312(5781), 1780-1782.
19. Leonhardt, U. (2006). *Optical conformal mapping*. Science, 312(5781), 1777-1780.
20. Lin Xu and Huanyang Chen (2015) *Conformal transformation optics*, Nature Photonics, 9, 15-23.
21. Ward, A. J., & Pendry, J. B. (1996). *Refraction and geometry in Maxwell's equations*. Journal of Modern Optics, 43(4), 773-793.
22. Leonhardt, U., & Philbin, T. G. (2009). *Transformation optics and the geometry of light*. Progress in Optics, 53(08), 69-152.
23. Thompson, R. T., Cummer, S. A., & Frauendiener, J. (2011). *Generalized transformation optics of linear materials*. Journal of Optics, 13(5), 055105.
24. Vakil A. and Engheta N. *Transformation Optics using Graphene*, Science 332, 1291-1294 (2011)
25. Martini, E., & Maci, S. (2013, April). *Transformation optics applied to metasurfaces*. In Antennas and Propagation (EuCAP), 2013 7th European Conference on (pp.1830-1831). IEEE.
26. Maci, S., Minatti, G., Casaletti, M., & Bosiljevac, M. (2011). *Metasurfaces: Addressing waves on impenetrable metasurfaces*. Antennas and Wireless Propagation Letters, IEEE, 10, 1499-1502.
27. Mencagli, M., Martini, E., Gonzalez-Ovejero, D., & Maci, S. (2014, April). *Transformation optics for anisotropic metasurfaces.* In Antennas and Propagation (EuCAP), 2014 8th European Conference on (pp. 1712-1716). IEEE.
28. Mencagli, M., Martini, E., Gonzlez-Ovejero, D., & Maci, S. (2014). *Metasurface transformation optics*. Journal of Optics, 16(12), 125106.
29. Lee, J.M. (2012). *Introduction to Smooth Manifolds*. Springer New York: Graduate Texts in Mathematics.
30. Frankel, T. (2004). *The Geometry of Physics: An Introduction*. Cambridge University Press.

31. Kuester, E. F., Mohamed, M. A., Piket-May, M., & Holloway, C. L. (2003). *Averaged transition conditions for electromagnetic fields at a metafilm*. Antennas and Propagation, IEEE Transactions on, 51(10), 2641-2651.
32. Holloway, C. L., Love, D. C., Kuester, E. F., Gordon, J. A., & Hill, D. A. (2012). *Use of generalized sheet transition conditions to model guided waves on metasurfaces/metafilms*. Antennas and Propagation, IEEE Transactions on, 60(11), 5173-5186.
33. Kuester, E. F., Holloway, C. L., & Mohamed, M. A. (2010, July). *A generalized sheet transition condition model for a metafilm partially embedded in an interface*. In Antennas and Propagation Society International Symposium (APSURSI), 2010 IEEE (pp. 1-4). IEEE.
34. Holloway, C. L., Kuester, E. F., Gordon, J. A., O'Hara, J., Booth, J., & Smith, D.R. (2012). *An overview of the theory and applications of metasurfaces: The two-dimensional equivalents of metamaterials*. Antennas and Propagation Magazine, IEEE, 54(2), 10-35.
35. Zhao, Y., Engheta, N., & Alu, A. (2011). *Homogenization of plasmonic metasurfaces modeled as transmission-line loads*. Metamaterials, 5(2), 90-96.
36. Idemen, M. (1973). *The Maxwell's equations in the sense of distributions*. Antennas and Propagation, IEEE Transactions on, 21(5), 736-738.
37. Idemen, M. (1990). *Universal boundary relations of the electromagnetic field*. Journal of the Physical Society of Japan, 59(1), 71-80.
38. Zhao, Y., Liu, X. X., & Alu, A. (2014*). Recent advances on optical metasurfaces*. Journal of Optics, 16(12), 123001.
39. Holloway, C. L., Kuester, E. F., & Dienstfrey, A. (2011). *Characterizing metasurfaces/metafilms: The connection between surface susceptibilities and effective material properties*. Antennas and Wireless Propagation Letters, IEEE, 10, 1507-1511.
40. Patel, A. M. (2013). *Controlling electromagnetic surface waves with scalar and tensor impedance surfaces* (Doctoral dissertation, Massachusetts Institute of Technology).
41. Achouri, K., Salem, M. A., & Caloz, C. (2014). *General metasurface synthesis based on susceptibility tensors*. arXiv preprint arXiv:1408.0273.
42. C. Pfeiffer A. Grbic Metamaterial *Huygens' Surfaces: Tailoring Wave Fronts with Reflectionless Sheets* Phys. Rev. Lett. 110, 197401 (2013)
43. Zheludev N. and Kivshar Y.S., *From Metamaterials to Metadevices*, Nature Materials, 11, 917-924 (2012)
44. Chen, H., Chan, T.C. and Sheng, P., *Transformation optics and metamaterials*, Nature Materials, 9, 387-369 (2010).
45. Duan Z., et al. *Reversed Cherenkov radiation in a waveguide filled with anisotropic double-negative metamaterials*. J. Appl. Phys., **104** (2008): 063303
46. Smith, D.R., Pendry, J.B. and Wiltshire, M.C.K., *Metamaterials and Negative Refractive Index*, Science, 305, 788-792 (2004).
47. D. R. Smith and J. B. Pendry "Homogenization of metamaterials by field averaging". JOSA B, 23, 391 (2005).
48. Kumar, A., Fung, K. H., Homer Reid, M. T., & Fang, N. X. (2014). *Transformation optics scheme for two-dimensional materials*. Optics letters, 39(7), 2113-2116.

**Acknowledgments**

We gratefully acknowledge partial financial support from *the Science and Engineering Research Council* (*SERC*) under grant numbers 1426500053 and 1426500054.

**Supplementary Information** is linked to the online version of the paper.

**Author Contributions**

J.T.Y.H., L.J.W., C.M., and P.G. developed the theoretical descriptions. P.G. directed the research. All authors contributed to writing the manuscript.

**Competing Financial Interests**

The authors declare no competing financial interests.